\documentclass[12pt,dvips]{article}
\usepackage{rotating}
\usepackage{hhline}
\usepackage{array}
\usepackage{epsfig}

\advance\textheight by \topskip
\begin{document}
\tolerance=100000
\thispagestyle{empty}
\setcounter{page}{0}
\newcommand{\eeHZ}{$e^+e^- \to ZH$}
\newcommand{\epem}{$e^+e^-$}

\newcommand{\nn}{\nonumber}
\newcommand{\be}{\begin{equation}}
\newcommand{\ee}{\end{equation}}
\newcommand{\ba}{\begin{eqnarray}}
\newcommand{\ea}{\end{eqnarray}}
\newcommand{\bann}{\begin{eqnarray*}}
\newcommand{\eann}{\end{eqnarray*}}
\newcommand{\bc}{\begin{center}}
\newcommand{\ec}{\end{center}}
\newcommand{\SM}{{\sc Sm~}}
\newcommand{\SUSY}{{\sc Susy~}}
\newcommand{\MSSM}{{\sc Mssm~}}
\newcommand{\LEP}{{\sc Lep}}
\newcommand{\LHC}{{\sc Lhc}}
\newcommand{\sts}{\scriptstyle}
\newcommand{\ngs}{\!\!\!\!\!\!}
\newcommand{\rb}[2]{\raisebox{#1}[-#1]{#2}}
\newcommand{\CP}{${\cal CP}$~}

\begin{titlepage}

\begin{flushright}
DESY 01-010\\
PM/01-08\\
LC-TH-2001-033\\
\end{flushright}

\vspace{1cm}

\renewcommand{\thefootnote}{\fnsymbol{footnote}}
\begin{center}
  {\Large \bf Measuring the Spin of the Higgs Boson\footnote[1]{Supported in part by the European Union (HPRN-CT-2000-00149) and by the Korean Research Foundation (KRF-2000-015-050009).}}\\[1cm]
{\large D.J.~Miller$^1$, S.Y.~Choi$^{1,\,2}$, B.~Eberle$^1$, M.M.~M\"uhlleitner$^{1, \, 3}$ \\ 
and P.M.~Zerwas$^{1}$\\[1cm]
 {\it 
$^1$ Deutsches Elektronen--Synchrotron DESY, D--22603 Hamburg, Germany\\
$^2$ Chonbuk National University, Chonju 561-756, Korea\\
$^3$ Universit\'e de Montpellier II, F--34095 Montpellier Cedex 5, France}}
\end{center}

\renewcommand{\thefootnote}{\arabic{footnote}}
\vspace{3cm}

\begin{abstract}
\noindent
By studying the threshold dependence of the excitation curve and the
angular distribution in Higgs-strahlung at \epem colliders, \eeHZ, the
spin of the Higgs boson in the Standard Model and related extensions
can be determined unambiguously in a model-independent way.
\end{abstract}

\end{titlepage}

\noindent
{\bf 1.} Establishing the Higgs mechanism for generating the masses of
the fundamental particles, leptons, quarks and gauge bosons, in the
Standard Model and related extensions, is one of the principal aims of
experiments at prospective \epem~linear
colliders~\cite{linear_collider}. After the experimental clarification
of tantalizing indications of a light Higgs boson at
\LEP~\cite{Light_Higgs} has been stopped, the particle can be discovered
at the Tevatron~\cite{Tevatron} or later at the \LHC~\cite{LHC}.

Assuming the positive outcome of these experiments, we address in this
letter the question of how the spinless nature and the positive parity
of the Higgs boson\footnote{The determination of the parity and the
  parity mixing of a spinless Higgs boson has been extensively
  investigated in Refs.~\cite{Kramer,haggun}.} can be established in a
model independent way.  Higgs-strahlung, \be e^+e^- \to ZH, \ee
provides the mechanism for the solution of this problem.  The rise of
the excitation curve near the threshold and the angular distributions
render the spin-parity analysis of the Higgs boson unambiguous in this
channel.  Without loss of generality, we can assume the Higgs boson to
be emitted from the $Z$-boson line, Fig.~\ref{fig:fd}(a). Were it
emitted from the lepton line\footnote{We thank H.~Murayama and
  T.~Rizzo for alerting us to this potential loophole.}, the required
$Hee$ coupling would be so large that the state could have been
detected as a resonance at \LEP, $e^+e^-\to H(\gamma)$, or could be
detected at the \LHC~ via resonant $H \to e^+e^-$ decays, dominating
over the $H \to ZZ^{(*)} \to 4l$ decay mode which involves two small
$Z$ branching ratios.

The cross section for Higgs-strahlung in the Standard Model is given
by the expression~\cite{Higgsstrahlung} 
\be \sigma[e^+e^- \to ZH] =
\frac{G_F^2 M_Z^4}{96 \pi s} \, (v_e^2+a_e^2) \, \beta \,
\frac{\beta^2+12M_Z^2/s}{(1-M_Z^2/s)^2}, \ee 
where $v_e=-1+4\sin^2 \theta_W$ and $a_e=-1$ are the vector and
axial-vector $Z$ charges of the electron; $M_Z$ is the $Z$-boson mass,
$\sqrt{s}$ the centre-of-mass energy, and $\beta=2p/\sqrt{s}$ the
$Z$/$H$ three-momentum in the centre-of-mass frame, in units of the
beam energy, \mbox{{\it i.e.} $\beta^2
  =[1-(M_H+M_Z)^2/s][1-(M_H-M_Z)^2/s]$}.  The excitation curve rises
linearly with $\beta$ and therefore steeply with the energy above the
threshold\footnote{Non-zero width effects can easily be incorporated;
  see A. Para, note in preparation.}: \be \sigma \sim \beta \sim
\sqrt{s-(M_H+M_Z)^2}. \ee This rise is characteristic of the
production of a scalar particle in conjunction with the $Z$ boson
(with only two exceptions, to be discussed later). 

The second characteristic is the angular distribution of the Higgs and
$Z$ bosons in the final state~\cite{Ang_dist},
\be \frac{1}{\sigma}
\frac{d\sigma}{d \cos \theta} = \frac{3}{4} \frac{\beta^2 \sin^2
  \theta +8 M_Z^2/s}{\beta^2+12 M_Z^2/s}.
\label{eq:dsm} \ee
The distribution of the polar angle $\theta$ is isotropic near the
threshold and it develops into the characteristic $\sin^2 \theta$ law at
high energies which corresponds to dominant longitudinal $Z$ production, 
congruent with the equivalence theorem.

Independent information on the helicity of the $Z$ state is encoded in
the final-state fermion distributions in the decay $Z \to f \bar f$.
Denoting the fermion polar angle\footnote{Azimuthal distributions
  provide supplementary information, see Ref.~\cite{Ang_dist}; to
  match the definitions used in the formulae, the azimuthal angle shown
  in Fig.~9(a) of Ref.~\cite{Ang_dist} should be denoted $(\pi -
  \phi_*)$.} in the $Z$ rest frame with respect to the $Z$ flight
direction in the laboratory frame by $\theta_*$, the double
differential distribution in $\theta$ and $\theta_*$ is predicted by
the Standard Model to be
\ba \frac{1}{\sigma} \frac{d\sigma}{d\cos \theta \, d\cos\theta_*} 
&=& \frac{9M_Z^2 \gamma^2 /4s}{\beta^2+12M_Z^2/s} 
\bigg\{ \sin^2\theta \sin^2\theta_* 
+ \frac{1}{2\gamma^2}[1+\cos^2\theta][1+\cos^2\theta_*] \nn \\
&&\phantom{\frac{9M_Z^2/8s}{\beta^2+12M_Z^2/s} \bigg\{}
+ \frac{1}{2\gamma^2} \frac{2 \, v_ea_e}{(v_e^2+a_e^2)}
\frac{2\,v_f a_f}{(v_f^2+a_f^2)}\,4 \cos\theta \cos\theta_* \bigg\}, \ea
with $\gamma^2=E_Z^2/M_Z^2=1+\beta^2s/4M_Z^2$. Again, for high energies, 
the longitudinal $Z$ polarization is reflected in the asymptotic behaviour 
$\propto \sin^2\theta_*$. \\

\noindent
{\bf 2.} The helicity formalism is the most convenient theoretical
tool for defining observables which uniquely prove the scalar nature
of the Standard-Model Higgs boson. Denoting the basic helicity
amplitude~\cite{Helicity} for arbitrary $H$ spin-${\cal J}$, with the 
azimuthal angle set to zero, by 
\be \langle
Z(\lambda_Z) H(\lambda_H) | Z^*(m) \rangle = \frac{g_W M_Z}{\cos
  \theta_W} \, d^1_{m,\,\lambda_Z-\lambda_H}(\theta) \, \Gamma_{\lambda_Z
  \lambda_H}, \label{eq:vert} \ee 
the reduced vertex $\Gamma_{\lambda_Z \lambda_H}$ is dependent only on
the helicities of the $Z$ and Higgs bosons, $\lambda_Z$ and
$\lambda_H$ respectively, and is independent of the $Z^*$ spin
component $m$ along the beam-axis by rotational invariance. The
standard coupling is split off explicitly.

The normality of the Higgs state, \be n_H = (-1)^{\cal J} \,{\cal P}, \ee
which is the product of the spin signature $(-1)^{\cal J}$ and the parity
${\cal P}$, plays an important r\^ole in classifying these helicity
amplitudes. The normality determines the relation between helicity
amplitudes under parity transformations. If the interactions which
determine the vertex (\ref{eq:vert}) are ${\cal P}$ invariant,
equivalent to \CP invariance in this specific case, the reduced
vertices are related by
\be \Gamma_{\lambda_Z \lambda_H} = n_H \, \Gamma_{-\lambda_Z \,
  -\lambda_H}. \label{eq:norm} \ee 

The total cross section for a \CP invariant theory is in this
formalism then given by,
\be \sigma = \frac{G_F^2 M_Z^6\, (v_e^2+a_e^2)}{24 \pi s^2 \,(1-M_Z^2/s)^2} 
\, \beta \left[ |\Gamma_{00}|^2 + 2 \, |\Gamma_{11}|^2
  + 2 \, |\Gamma_{01}|^2 + 2 \, |\Gamma_{10}|^2 + 2 \, |\Gamma_{12}|^2
\right], \label{eq:hel} \ee
Correspondingly, the polar angular distributions introduced above can
be written,
\be \frac{1}{\sigma} \, \frac{d\sigma}{d\cos\theta} 
= \frac{3}{4 \, \Gamma^2}
\left\{ \sin^2 \theta
\left [|\Gamma_{00}|^2+2\,|\Gamma_{11}|^2 \right]  + [1+\cos^2\theta] 
\left[|\Gamma_{01}|^2+|\Gamma_{10}|^2+|\Gamma_{12}|^2 \right] \right\},
\label{eq:dhel} \ee
and
\ba \frac{1}{\sigma} \, \frac{d\sigma}{d\cos\theta \, d\cos\theta_*} 
&=& \frac{9}{16 \, \Gamma^2} 
\bigg\{ \sin^2\theta \sin^2\theta_* \, |\Gamma_{00}|^2 
+ \frac{1}{2}[1+\cos^2\theta][1+\cos^2\theta_*] \left[ |\Gamma_{10}|^2 
+ |\Gamma_{12}|^2 \right] \nn \\
&&  \phantom{\frac{9}{16 \, \Gamma^2} \bigg\{} 
+\sin^2\theta \, [1+\cos^2\theta_*] |\Gamma_{11}|^2 
+ [1+\cos^2\theta] \, \sin^2\theta_* |\Gamma_{01}|^2 \nn \\
&&  \phantom{\frac{9}{16 \, \Gamma^2} \bigg\{} 
+\frac{2\,v_ea_e}{(v_e^2+a_e^2)} \frac{2\,v_fa_f}{(v_f^2+a_f^2)}
\, 2 \cos\theta \cos\theta_* \left[ |\Gamma_{10}|^2-|\Gamma_{12}|^2 \right] 
\bigg\},
\label{eq:ddhel} \ea
where $\Gamma^2$ corresponds to the square bracket of Eq.~(\ref{eq:hel}).

The helicity amplitudes of Higgs-strahlung in the Standard Model are given by
\ba
\Gamma_{00}&=&-E_Z/M_Z, \nn \\
\Gamma_{10}&=&-1, \nn \\
\Gamma_{01}&=&\Gamma_{11} \,\, = \,\, \Gamma_{12} \,\, = \,\, 0,  \ea
and the Higgs boson carries even normality: \be n_H =+1. \ee

These amplitudes determine uniquely the spin-parity quantum numbers of
the Higgs boson; this will be demonstrated for a \CP invariant
theory, for even and odd normality Higgs bosons in {\bf 3a} and {\bf
  3b} respectively.  The analysis will be extended to mixed parity
assignments in \CP noninvariant theories thereafter.\\

{\bf 3a.} States of \underline{even normality} ${\cal J^P}=1^-$, $2^+$, $3^-$
\ldots. can be excluded by measuring the threshold behaviour of the
excitation curve and the angular correlations\footnote{It is well known
that the observation of $H \to \gamma \gamma$ decays or the formation of Higgs
bosons, $\gamma \gamma \to H$, in photon collisions rules out the spin-1
assignment as a result of the Landau-Yang theorem.}.

The most general current describing the $Z^*ZH$ vertex in
Fig.~\ref{fig:fd}(a) is given by the expression
\be {\cal J}_{\mu} = \frac{g_W M_Z}{\cos \theta_W} T_{\mu \alpha \beta_1 ... \beta_S} \, \varepsilon^*(Z)^{\alpha} \, \varepsilon^*(H)^{\beta_1 ... \beta_{\cal J}}. \label{eq:cur} \ee
While $\varepsilon^{\alpha}$ is the usual spin-1 polarization vector,
the spin-${\cal J}$ polarization tensor $\varepsilon^{\beta_1 ...
  \beta_{\cal J}}$ of the state $H$ has the notable properties of
being symmetric, traceless and orthogonal to the 4-momentum of the
Higgs boson $p_H^{\beta_i}$, and can be constructed from products of
suitably chosen polarization vectors. Moreover $T_{\mu \alpha \beta_1
  ... \beta_{\cal J}}$, normalized such that $T_{\mu \,
  \alpha}=g_{\perp \, \mu \, \alpha}$ in the Standard Model, is
transverse due to the conservation of the lepton current. These
properties strongly constrain the form of the tensor.  The most
general tensor for spins~$\leq 2$ can be seen in Tab.\ref{tab:ff}(top)
together with the resulting helicity amplitudes.  (The coefficients
$a_i$, $b_i$ and $c_i$ in Tab.\ref{tab:ff} are independent of the
momenta near the threshold.) The leading $\beta$ dependence of the
helicity amplitudes can be predicted from the form of the $Z^*ZH$
coupling.  Each momentum contracted with the $Z$-boson polarization
vector or the $H$ polarization tensor will necessarily give zero or
one power of $\beta$:
\be p_i \cdot \varepsilon_j(\lambda_j) = \left\{ 
\begin{array}{ll}
\beta s/2M_j  & \mathrm{ for } \quad i \neq j \quad \mathrm{ and } \quad \lambda_j=0 \\
0 & \mathrm{ for } \quad i=j=Z/H \quad \mathrm{ or } \quad \lambda_j=\pm. \end{array} \right.
\ee
Furthermore, any momentum contracted with the lepton current will also
give rise to one power of $\beta$ due to the transversality of the
current. Then, one need only count the number of momenta in each term
of $T^{\mu \alpha \beta_1 ... \beta_{\cal J}}$ to understand the threshold
behaviour of the corresponding helicity amplitudes. The $\beta$
dependence of the excitation curve can be derived from the squared
$\beta$ dependence of the helicity amplitude multiplied by a single
factor $\beta$ from the phase space.

\underline{\bf Spin 0:} The spin-0 helicity amplitudes presented in
Tab.{\ref{tab:ff}(top) have no dependence on $\beta$ near threshold.
  Consequently the excitation curve rises linearly in $\beta$ at
  threshold, with the single power of $\beta$ coming from the phase
  space. This is also the case for the Standard Model, as described in
  {\bf 1} and obtained from the spin-0 form factors by setting
  $a_1=1$ and $a_2=0$.

\underline{\bf Spin 1:} It is easily seen that all helicity amplitudes
vanish near threshold linearly in $\beta$, so the excitation curve
rises $\sim \beta^3$, distinct from the Standard Model.

\underline{\bf Spin 2:} The most general spin-2 tensor contains a term
with no momentum dependence ($\propto c_1$), resulting in helicity
amplitudes which do not vanish at threshold if $c_1 \neq 0$.  However,
the helicity amplitudes $\Gamma_{01}$ and $\Gamma_{11}$ contain
$c_1$ and are consequently non-zero in this case,
leading to non-trivial $[1+\cos^2\theta] \sin^2 \theta_*$ and $\sin^2
\theta [1+\cos^2\theta_*]$ correlations which are absent in the
Standard Model.  Therefore, if the excitation curve rises linearly,
not observing these correlations in experiment rules out the spin-2
assignment to the $H$ state. However, if $c_1=0$ in the spin-$2$
case, the excitation curve rises $\sim \beta^5$ near threshold.

\underline{\bf Spin $\geq 3$:} Above spin-2 the number of independent
helicity amplitudes does not increase any more~\cite{Helicity}.
Consequently, the most general spin-${\cal J}$ tensor $T_{\mu \alpha \beta_1
  ... \beta_{\cal J}}$ is a direct product of a tensor $T^{\mu \alpha \beta_i
  \beta_j}_{(2)}$ isomorphic with the spin-2 tensor and momentum
vectors \mbox{$q^{\beta_k}=(p_Z+p_H)^{\beta_k}$} as required by the
properties of the spin-${\cal J}$ wave-function $\varepsilon^{\beta_1 ...
  \beta_{\cal J}}$,
\be T^{\mu \alpha \beta_1 \beta_2 ... \beta_{\cal J}}_{({\cal J})} = \sum_{i<j}
T^{\mu \alpha \beta_i \beta_j}_{(2)} q^{\beta_1}...q^{\beta_{i-1}}
q^{\beta_{i+1}}...q^{\beta_{j-1}} q^{\beta_{j+1}}...q^{\beta_{\cal J}}.
\label{eq:hst} \ee
Contracted with the wave-function, the extra ${\cal J}-2$ momenta give rise
to a leading power $\beta^{{\cal J}-2}$ in the helicity amplitudes. The
cross section therefore rises near threshold $\sim \beta^{2{\cal J}-3}$, 
{\it i.e.} with a power $\geq 3$, in contrast to  the single power of 
the Standard Model. \\[0.1cm]

{\bf 3b.} It is quite easy to rule out particles of \underline{odd
  normality}, ${\cal J^P}=0^-$, $1^+$, $2^-$, \ldots, which may mimic
the Standard Model Higgs boson in Higgs-strahlung. Since the helicity
amplitude $\Gamma_{00}$ must vanish by Eq.~(\ref{eq:norm}), the
observation of a non-zero $\sin^2\theta \, \sin^2 \theta_*$
correlation in Eq.~(\ref{eq:ddhel}), as predicted by the Standard
Model, eliminates all odd normality states. In particular, the
assignment of negative parity to the spin-0 state can be ruled out by
observing~\cite{Privcom} a polar-angle distribution different from the
energy-independent $[1+\cos^2\theta]$ distribution which is
characteristic for $0^-$ particle production~\cite{Ang_dist} in
contrast to the Standard Model.

Nevertheless, in anticipation of the mixed normality scenario we
present the helicity amplitudes also for Higgs bosons of odd
normality, and spin~$\leq 2$ in Tab.\ref{tab:ff}(bottom). We find a
similar picture to the even normality case, where here the excitation
curve only presents a linear rise for a particle of spin-1. The
generalization to higher spins $\geq 3$ follows exactly as before,
resulting in an excitation curve $\sim \beta^{2{\cal J}-1}$, {\it i.e.} with 
a power $\geq 5$, at threshold.\\[0.1cm]

The above formalism can be generalized easily to rule out a mixed
normality state with spin~$\geq 1$. For a Higgs boson of mixed
normality one may no longer use Eq.~(\ref{eq:norm}) to obtain the
simple form of the (differential) cross sections seen in
Eqs.~(\ref{eq:hel}--\ref{eq:ddhel}). In particular, the polar angle
distribution, Eq.~(\ref{eq:dhel}), is modified to include a linear
term proportional to $\cos\theta$, indicative of 
\CP violation~\cite{haggun}.  The analysis, however, proceeds as in
the fixed normality case, since the most general tensor vertex will be
the sum of the even and odd normality tensors given in
Tab.\ref{tab:ff}.

A mixed normality Higgs boson of spin~$\geq 3$ may be eliminated by a
non-linear rise of the excitation curve at threshold, whereas those of
spin-1 and spin-2 may exhibit a linear $\beta$ dependence, arising
from the odd and even tensor contributions respectively. However,
these two exceptions can be ruled out by observing neither 
$[1+\cos^2\theta] \sin^2 \theta_*$ nor 
$\sin^2 \theta [1+\cos^2\theta_*]$ angular correlations, since a 
linear excitation curve in both cases requires that both $\Gamma_{01}$ 
and $\Gamma_{11}$ be non-zero.\\[0.1cm]


{\bf 4.} The analyses described above, can be summarized in a few
characteristic observations. The key is the threshold behaviour of the
excitation curve which is predicted to be linear in $\beta$ for the
${\cal J^P}=0^+$ Higgs boson within the Standard Model.  The
observation of the linear rise, if supplemented by the angular
correlations for two exceptional cases, rules out all other 
${\cal J^P}$ assignments:\\ \vspace{-0.5cm}
\begin{center}
\begin{tabular}{|lcl|} \hline
\rb{-0.3mm}{$\sigma \sim \sqrt{s-(M_Z+M_H)^2}\,\,$}
& (i) $\!\!\!$ & rules out ${\cal J^P}=0^-,1^-, 2^-, 3^{\pm}, ...$ \\
threshold:& (ii) $\!\!\!$& rules out ${\cal J^P}= \;\;\;\;\;\,1^+, 2^+$\\
&& if no $[1+\cos^2 \theta]\sin^2\theta_*$ 
$\sin^2\theta[1+\cos^2 \theta_*]$ correlations \\\hline
\end{tabular} \end{center} \vspace{0.1cm}

\noindent The same rules also eliminate all spin states ${\cal J}\geq1$ 
for mixed-normality assignments.

The rules can be supplemented by other observables which are specific
to two interesting cases. By observing a non-zero $H\gamma\gamma$
coupling, the spin-$1$ assignment can be ruled out independently.
Moreover, the negative-parity assignment in the spin-$0$ case would
give rise to the energy-independent angular distribution $\sim
[1+\cos^2 \theta]$ in contrast to scalar Higgs production, while mixed
\CP noninvariant $0^\pm$ assignments can be probed in a linear
$\cos\theta$ dependence of the Higgs-strahlung cross section.

As a result, the measurement of the threshold behaviour of the
excitation curve for Higgs-strahlung combined with angular
correlations can be used to establish the ${\cal J^P}=0^+$ character
of the Higgs boson in the Standard Model and related extensions
unambiguously.

\section*{Acknowledgments}

Thanks go to D.J.~Miller for continual encouragement during the
project. We are grateful to K.~Desch and A.~Para for useful experimental
advice, and to G.~Kramer for discussions and the critical reading of
the manuscript.

\begin{table}[h]
\centering
\begin{tabular}{||c||l|l|c||} \hhline{|t:=:t:===:t|}
${\cal J^P}$&\multicolumn{1}{c|}{\small $Z^*ZH$ Coupling}
& \multicolumn{1}{c|}{\small Helicity Amplitudes} & \small Threshold
\\\hhline{|:=:b:===:|} 
\multicolumn{4}{||c||}{\small Even Normality $n_H=+$} \\\hhline{|:=:t:===:|}
&&$\sts \Gamma_{00} =  (-a_1E_Z-\frac{1}{2}\,a_2\,s^{3/2}\, \beta^2)/M_Z$ & $\sts 1$ \\
 \rb{1.5ex}{$0^+$} 
&\rb{1.5ex}{$\sts \phantom{+}a_1 g^{\mu\alpha}_{\perp}+a_2 k^{\mu}_{\perp}q^{\alpha}$}
& $\sts \Gamma_{10} = -a_1$ 
& $\sts 1$ \\\hhline{||-||---||}
&&$\sts \Gamma_{00}= \beta \, [-b_1\,(s-M_Z^2-M_H^2)-\frac{1}{2}\,b_2\,s^2\,\beta^2+b_3\,s$&\\
&$\sts \phantom{+}b_1 \, g^{\alpha\beta} k^{\mu}_{\perp}+b_2 \, q^{\alpha}q^{\beta} k^{\mu}_{\perp}$
&$\sts \phantom{\Gamma_{00}=}\quad-b_4\,(M_Z^2-M^2_H)]\sqrt{s}/(2\,M_Z M_H)$ 
& \rb{1.5ex}{$\sts \beta$} \\
$1^-$
&$\sts +b_3 \, (q^{\alpha}g^{\mu\beta}_{\perp}-q^{\beta}g^{\mu\alpha}_{\perp})$
&$\sts \Gamma_{10}=\beta\,(b_3-b_4)\,s/(2\,M_H)$ & $\sts \beta$ \\
&$\sts +b_4 \, (q^{\alpha} g^{\mu\beta}_{\perp}+ q^{\beta} g^{\mu\alpha}_{\perp})$
&$\sts \Gamma_{01}=\beta\,(b_3+b_4)\,s/(2\,M_Z)$ & $\sts \beta$\\
&&$\sts \Gamma_{11}= \beta\,\sqrt{s}\,\,b_1 $ & $\sts \beta$ \\\hhline{||-||---||}
&&$\sts \Gamma_{00}=\frac{\sqrt{2/3}}{M_ZM_H^2}\,\big\{
c_1\,E_H(s-M_Z^2-M_H^2) -\frac{1}{8}c_5\,s^{7/2}\beta^4 $ &\\
&\rb{1.5ex}{$\sts \phantom{+}c_1\, (g^{\alpha\beta_1}g^{\mu\beta_2}_{\perp}
+g^{\alpha\beta_2}g^{\mu\beta_1}_{\perp})$}
&$\sts \phantom{\Gamma_{00}=}-\frac{1}{4} s^2 \beta^2[c_2\,E_Z-2\,c_3\,E_H
+2\,c_4\,(s-M_Z^2-M_H^2)/\sqrt{s} \, ] \big\}$ &\rb{1.5ex}{$\sts 1$}\\
&\rb{1.5ex}{$\sts +c_2\,g^{\mu\alpha}_{\perp}\,q^{\beta_1}q^{\beta_2}$}
&$\sts \Gamma_{10}=\sqrt{2/3}(-c_1-c_2s^2\beta^2/(4M_H^2))$ &$\sts 1$\\
\rb{1.5ex}{$2^+$}
&\rb{1.5ex}{$\sts +c_3\,(g^{\mu\beta_1}_{\perp}q^{\beta_2}+g^{\mu\beta_2}_{\perp}q^{\beta_1})\,q^{\alpha}$}
&$\sts \Gamma_{01}=(2\,c_1(s-M_Z^2-M_H^2)+c_3\,s^2\beta^2)/(2\sqrt{2}M_ZM_H)$&$\sts 1$\\
&\rb{1.5ex}{$\sts +c_4\,(g^{\alpha\beta_1} q^{\beta_2}+g^{\alpha\beta_2} q^{\beta_1}) k^{\mu}_{\perp}$}
&$\sts \Gamma_{11}=(-c_1\,E_H+\frac{1}{2}\,c_4\,s^{3/2}\beta^2)\sqrt{2}/M_H$&$\sts 1$\\
&\rb{1.5ex}{$\sts +c_5\,k^{\mu}_{\perp}q^{\alpha}q^{\beta_1}q^{\beta_2}$}
&$\sts \Gamma_{12}=-2\, c_1$&$\sts 1$\\\hhline{|:=:b:===:|} 
\multicolumn{4}{||c||}{\small Odd Normality $n_H=-$} \\\hhline{|:=:t:===:|}
&&$\sts\Gamma_{00}=0$&\\
\rb{1.5ex}{$0^-$}&
\rb{1.5ex}{$\sts \phantom{+}a_1\,\epsilon^{\mu\alpha\rho\sigma}q_{\rho}k_{\sigma}$}
&$\sts\Gamma_{10}=-i\,\beta\,s\,a_1$&$\sts \beta$\\\hhline{||-||---||}
&&$\sts\Gamma_{00}=0$&\\
&\rb{1.5ex}{$\sts\phantom{+}b_1\,\epsilon^{\mu\alpha\beta\rho}q_{\rho}$}
&$\sts\Gamma_{10}=-i\,(b_1\,s\,E_H+b_2\,(E_H\,(M_Z^2-M_H^2)+\frac{1}{2}\,s^{3/2}\beta^2))/(\sqrt{s}\,M_H)$
&$\sts 1$\\
\rb{1.5ex}{$1^+$}
&\rb{1.5ex}{$\sts +b_2\,\epsilon^{\mu\alpha\beta\rho}_{\perp}k_{\rho}$}
&$\sts\Gamma_{01}=-i\,(b_1\,s\,E_Z+b_2\,(E_Z\,(M_Z^2-M_H^2)-\frac{1}{2}\,s^{3/2}\beta^2))/(\sqrt{s}\,M_Z)$
&$\sts 1$\\
&\rb{1.5ex}{$\sts+b_3\,\epsilon^{\alpha\beta\rho\sigma}q_{\rho}k_{\sigma}k^{\mu}_{\perp}$}
&$\sts\Gamma_{11}=-i\,(b_1\,s+b_2\,(M_Z^2-M_H^2)+b_3\,s^2\beta^2)/\sqrt{s}$
&$\sts 1$\\\hhline{||-||---||}
&&$\sts\Gamma_{00}=0$&\\
&$\sts\phantom{+}c_1\,\epsilon^{\mu\alpha\beta_1\rho}q^{\beta_2}q_{\rho}$ 
&$\sts\Gamma_{10}=-i\,\beta\,(c_1\,sE_H+c_2\,(E_H(M_Z^2-M_H^2)+\frac{1}{2}s^{3/2}\beta^2)$
&\\
&$\sts+c_2\,\epsilon^{\mu\alpha\beta_1\rho}_{\perp}k_{\rho}q^{\beta_2}$
&$\sts\phantom{\Gamma_{10}=\beta\,(}+\frac{1}{4}\,c_4\,s^{5/2}\beta^2)\sqrt{2s}/(\sqrt{3}M_H^2)$
&\rb{1.5ex}{$\sts \beta$}\\
\rb{1.5ex}{$2^-$}
&$\sts+c_3\,\epsilon^{\alpha\beta_1\rho\sigma}q^{\beta_2}
k^{\mu}_{\perp}\,q_{\rho}k_{\sigma}$
&$\sts\Gamma_{01}=-i\,\beta\,(c_1\,sE_Z+c_2\,(E_Z(M_Z^2-M_H^2)$&\\
&$\sts+c_4 \, \frac{1}{2}\epsilon^{\mu\alpha\rho\sigma}q_{\rho}k_{\sigma} 
q^{\beta_1}q^{\beta_2}$ 
&$\sts\phantom{\Gamma_{01}=\beta\,(}-\frac{1}{2}s^{3/2}\beta^2))\sqrt{s}/(\sqrt{2}\,M_ZM_H)$&\rb{1.5ex}{$\sts \beta$}\\
&$\sts+ \beta_1 \leftrightarrow \beta_2$ 
&$\sts\Gamma_{11}=-i\,\beta\,(c_1\,s+c_2\,(M_Z^2-M_H^2)+c_3\,s^2\beta^2)\sqrt{s}/(\sqrt{2}\,M_H)$
&$\sts \beta$\\
&&$\sts\Gamma_{12}=0$& \\\hhline{|b:=:b:===:b|}
\end{tabular}
\caption{\it The most general tensor couplings of the $Z^*ZH$ vertex and the 
corresponding helicity amplitudes for Higgs bosons of spin~$\leq 2$. 
Here $q=p_Z+p_H$, \mbox{$k=p_Z-p_H$} and $_{\perp}$ indicates 
orthogonality of a vector or tensor to $q^{\mu}$, 
\mbox{$t_{\perp}^{\mu...}=t^{\mu...}-q^{\mu}/s\,q_{\nu}t^{\nu...}$}. 
For spin $\geq 3$, the helicity amplitudes rise $\sim \beta^{{\cal J}-2}$ and 
$\sim \beta^{{\cal J}-1}$ for even and odd normalities respectively.}
\label{tab:ff}
\end{table}

\begin{figure}[t]
\centering\epsfig{file=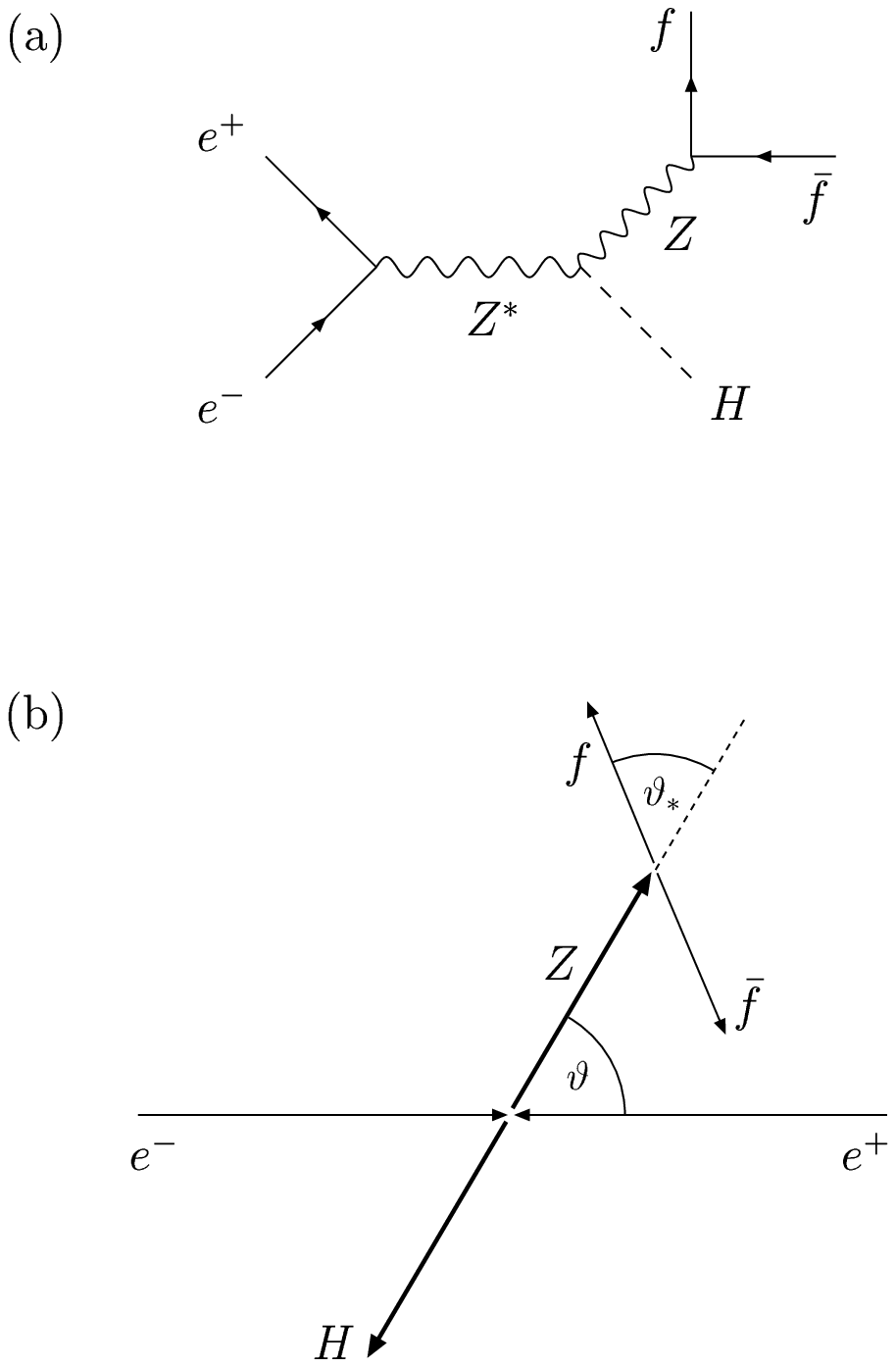} 
\caption{(a) The Higgs-strahlung process, $e^+e^- \to ZH$, followed by the subsequent  $Z$ boson decay $Z \to f \bar f$, and (b) the definition of the polar angles $\theta$ and $\theta_*$, for production and decay respectively.}
\label{fig:fd}
\end{figure}


\begin{thebibliography}{99}
  
\bibitem{linear_collider} H.~Murayama and M.E.~Peskin, Ann. Rev.
  Nucl. Part. Sci. {\bf 46} (1996) 533; E.~Accomando et al, Phys.
  Rept. {\bf 299} (1998) 1; P.~M.~Zerwas, Lectures, {\it Carg\`ese
    1999 Summer Institute}, Proceedings, [hep-ph/0003221].
  
\bibitem{Light_Higgs} R. Barate et al. (ALEPH Collaboration),
  Phys.~Lett. {\bf B495} (2000) 1; M. Acciarri et al. (L3
  Collaboration), Phys.~Lett. {\bf B495} (2000) 18; P.~Igo-Kemenes,
  LEPC presentation, Nov. 2000, {\tt
    http://lephiggs.web.cern.ch/LEPHIGGS/talks/index.html}.
  
\bibitem{Tevatron} M.~Carena et al, {\it Report of the Tevatron Higgs
    Working Group}, FERMILAB-CONF-00-279-T, [hep-ph/0010338].
  
\bibitem{LHC} ATLAS Collaboration, {\it Detector and physics
    performance Technical Design Report}, CERN-LHCC-99-14 \& 15
  (1999); CMS Collaboration, {\it Technical Design Report},
  CERN-LHCC-97-10 (1997).
  
\bibitem{Kramer} M.~Kr\"amer, J.~K\"uhn, M.~L.~Stong and P.~M.~Zerwas,
  Z.\ Phys.\ {\bf C64} (1994) 21.
  
\bibitem{haggun} K.~Hagiwara, S.~Ishihara, J.~Kamoshita and
  B.~A.~Kniehl, Eur.\ Phys.\ J.\ {\bf C14} (2000) 457; B.~Grzadkowski,
  J.F.~Gunion and J.~Pliszka, Nucl.  Phys. {\bf B583} (2000) 49;
  T.~Han and J.~Jiang, MADPH-00-1201, [hep-ph/0011271].

\bibitem{Higgsstrahlung}J.~Ellis, M.~K.~Gaillard and D.~V.~Nanopoulos,
  Nucl. Phys. {\bf B106} (1976) 292; B.~L.~Ioffe and V.~A.~Khoze, Sov.
  J. Part. Nucl. {\bf 9} (1978) 50; B.~W.~Lee, C.~Quigg and
  H.~B.~Thacker, Phys. Rev.  {\bf D16} (1977) 1519.
  
\bibitem{Ang_dist} V.~Barger, K.~Cheung, A.~Djouadi, B.~A.~Kniehl and
  P.~M.~Zerwas, Phys. Rev. {\bf D49} (1994) 79.
  
\bibitem{Helicity} G.~Kramer and T.~F.~Walsh, Z. Physik {\bf 263}
  (1973) 361.
  
\bibitem{Privcom} M.~Schumacher, LC-PHSM-2001-003; A.~Para, private
  communication.

\bibitem{Birgit} B.~Eberle, Diploma thesis, University of Hamburg, 2001.
  
\end{thebibliography}
\end{document}